\pdfoutput=1
\documentclass[ASNA,twocolumn]{USG}
\usepackage{anyfontsize} %

\usepackage{colortbl}   % For coloring table lines
\usepackage{xcolor}     % For color definitions

\graphicspath{{./images/}}

\newcommand{\revision}[1]{\textcolor{black}{#1}}

\newcommand{\widefigure}[2]{%
  \IfFileExists{#1}{%
    \includegraphics[width=\textwidth]{#1}%
  }{%
    \fbox{\parbox[c][#2][c]{0.96\textwidth}{%
      \centering\textbf{Figure placeholder}\\[4pt]
      \texttt{\detokenize{#1}}%
    }}%
  }%
}

\specialissue{AE-CAI | CARE | OR 2.0 | PRiSM Joint MICCAI Workshop 2026}
\articletype{ORIGINAL ARTICLE}
\journal{Healthcare Technology Letters}
\volume{0}
\copyyear{2026}
\startpage{1}
\articledoi{}

\begin{document}

\title{Feasibility of Augmented Reality-Guided Robotic Ultrasound with Cone-Beam CT Integration for Spine Procedures}
\titlemark{GEOMETRY-COUPLED MULTIVIEW REGISTRATION}

\author[1,2]{Feng Li}[https://orcid.org/0000-0003-2717-8274]
\author[1,2]{Tianyu Song}
\author[1]{Felix Pabst}
\author[1,2]{Miruna-Alexandra Gafencu}
\author[1]{Yordanka Velikova}
\author[1,2]{Yuan Bi}
\author[1]{Ulrich Eck}
\author[1,2]{Nassir Navab}

\authormark{Feng Li \textsc{et al.}}
\titlemark{GEOMETRY-COUPLED MULTIVIEW REGISTRATION}

\address[1]{\orgdiv{Chair for Computer-Aided Medical Procedures and Augmented Reality, }\orgname{Technical University of Munich, }%
\orgaddress{\state{Munich, }\country{Germany}}}

\address[2]{\orgdiv{}\orgname{Munich Center for Machine Learning, }%
\orgaddress{\state{Munich, }\country{Germany}}}

\corres{Feng Li (\email{feng.li@tum.de})}

\contributed{Feng Li, Tianyu Song, and Felix Pabst contributed equally to this study.}

% \editor{\textbf{Academic Editor:} Alex ~|~ \textbf{Guest Editor:} Andrew A. Rooney}

% Author, affiliation, correspondence, funding, and acknowledgement fields
\fundingInfo{This work was supported in part by the State of Bavaria through the Bayerische Forschungsstiftung (BFS) under Grant AZ-1592-23-ForNeRo.
This work was also supported in part by the Multi-Scale Medical Robotics Center, AIR@InnoHK, Hong Kong.
Additional support was provided by the Zentrales Innovationsprogramm Mittelstand (ZIM) project ``LIEBHERZ'' under Grant No.~16KN112326.}
% are intentionally omitted for the initial double-blind submission.
\historydates{}


\abstract[ABSTRACT]{
Accurate needle placement in spine interventions is critical for effective pain management, yet it depends on reliable identification of anatomical landmarks and careful trajectory planning. Conventional imaging guidance often relies on both computed tomography (CT) and X-ray fluoroscopy, exposing patients and staff to high doses of radiation while providing limited real-time 3D feedback. We present an optical see-through augmented reality (OST-AR)–guided robotic  system for spine procedures that provides in situ visualization of spinal structures to support needle trajectory planning. We integrate a cone-beam CT (CBCT)-derived 3D spine model, which is co-registered with live ultrasound, enabling users to combine global anatomical context with local, real-time imaging. We evaluated the system in a phantom user study involving two representative spine procedures: facet joint injection and lumbar puncture. Sixteen participants performed insertions under two visualization conditions: conventional screen vs. AR. Results show that AR significantly reduces execution time and across-task placement error, while also improving usability, trust, and spatial understanding, and lowering cognitive workload. These findings demonstrate the feasibility of AR-guided robotic ultrasound for spine interventions, highlighting its potential to enhance accuracy, efficiency, and user experience in image-guided procedures.
}

\maketitle

\section{Introduction}\label{sec:Introduction}

\par
Image guidance for spinal injection procedures requires accurate identification of target levels and safe needle trajectory planning within complex, patient-specific anatomy. Among these procedures, lumbar puncture for neuraxial access and facet joint injection for pain management are most common~\cite{wu2016effectiveness}. Conventional landmark-based approaches can be unreliable, particularly in patients with obesity or degenerative spinal changes, leading to multiple insertion attempts and increased patient discomfort~\cite{evansa2015ultrasound}. The success of these procedures depends on accurate needle placement and trajectory planning, requiring identification of spinal landmarks, including the target vertebral level, the interspinous space, and the midline entry path.

\par
In current clinical practice, spinal injections are most commonly performed under fluoroscopy or computed tomography (CT). CT provides volumetric context but is typically preoperative and not real-time. Physicians must therefore mentally map the anatomical structures observed in the CT volume to the patient during the intervention. Fluoroscopy, in contrast, provides quasi–real-time imaging but remains a two-dimensional modality. For injection procedures, insertion depth and alignment are critical to ensure safety and efficacy. This places a high demand on the physician’s spatial reasoning, requiring mental mapping of 2D images to 3D anatomy in order to guide the needle to the correct target.

More recently, ultrasound (US) has been investigated as a radiation-free, portable alternative imaging modality for spinal interventions. When combined with robotic control, US systems can provide real-time 2D image feedback as well as compounded 3D volumes to enhance spatial perception~\cite{bi2024machine}. As early as 2014, Rasoulian~\emph{et al.} \cite{rasoulian2015ultrasound} evaluated the feasibility of robotic US (RUS) in spinal surgery. To improve vertebral level localization, Tirindelli~\emph{et al.} \cite{tirindelli2020force} proposed fusing US data with force feedback from the robot end-effector. Hardware advances, such as compliant probe holders, further improved safe, adaptable probe–patient interaction~\cite{wang2023compliant}. Esteban~\emph{et al.} further demonstrated a clinically integrated RUS workflow in the operating room (OR) that acquires ultrasound sweeps, compounds them into a 3D volume, and uses a calibrated needle guide for robot-aligned targeting~\cite{esteban2018robotic}. This work established that RUS can reduce reliance on fluoroscopy while fitting into routine OR practice, and it set a practical template for scan planning, volumetric compounding, and guided alignment in spinal injection procedures.

\begin{figure}[t]
    \centering
    \includegraphics[width=\columnwidth]{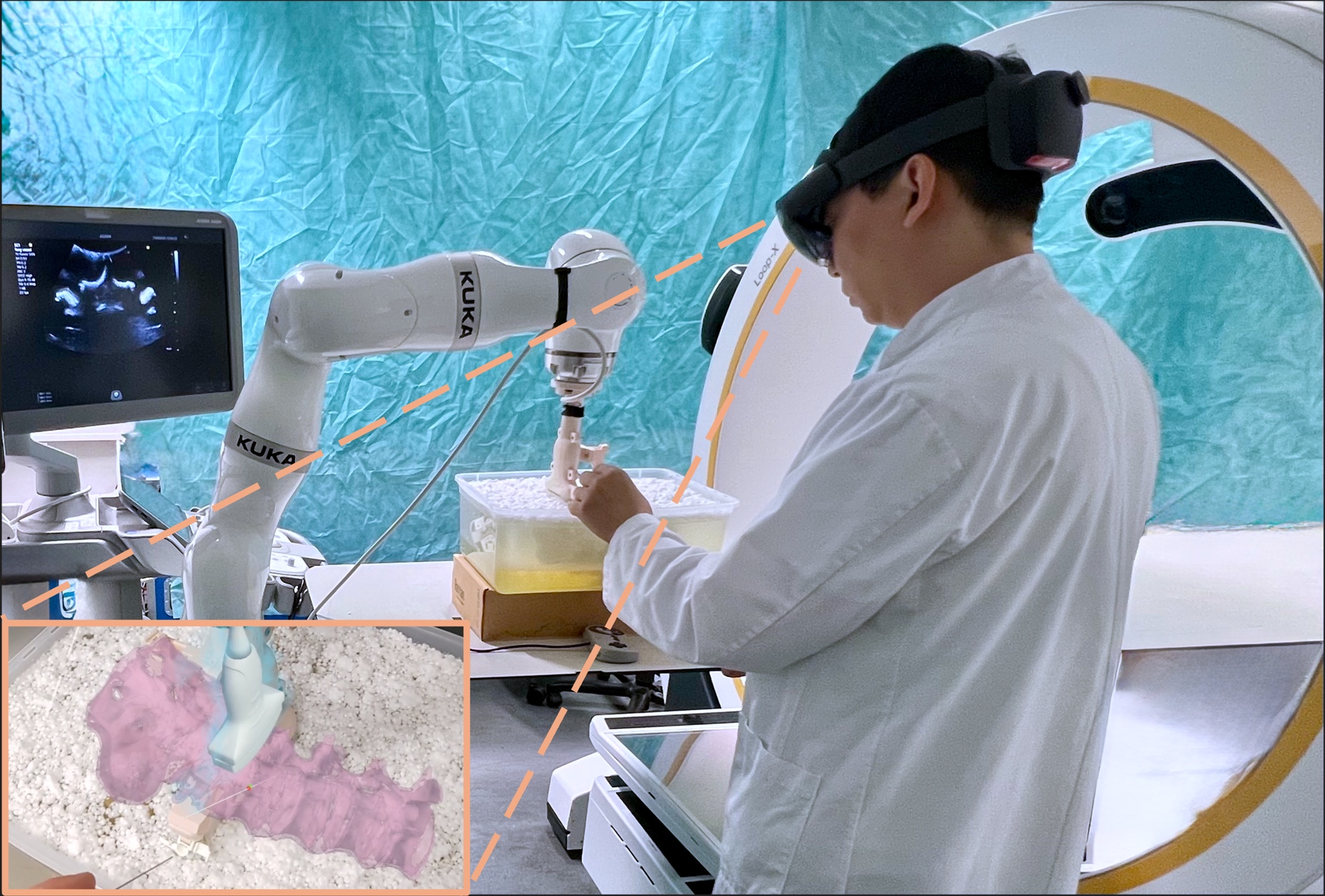}
    \caption{\textbf{AR-Guided Needle Insertion.} The system integrates robotic ultrasound, CBCT, and AR visualization into a unified framework, providing intuitive guidance during needle insertion for spine procedures.}
\label{fig:teaser}
\end{figure}

\par
Despite these advances, due to acoustic shadowing, US primarily reveals bone interfaces while deeper anatomy remains occluded, limiting global anatomical context during planning and guidance~\cite{gafencu2024shape}. Additionally, visualization is typically presented on distant screens, requiring clinicians to cognitively register global (compounded) and local (live) information to the patient at the bedside. To provide complete context, Li~\emph{et al.} \cite{li2025robotic} proposed a system that integrates robotic cone beam CT (CBCT) with  RUS. By co-registering the two modalities, the system provides a clear 3D visualization of the entire spine, while US offers real-time intraoperative feedback \cite{li2026ultrasound}. However, to enable accurate planning of the needle trajectory and effective real-time guidance during the procedure, an intuitive interaction platform and a streamlined workflow are required.

Augmented reality (AR) presents a promising solution by overlaying virtual information directly onto the real world, potentially bridging the gap between complex 3D anatomy and intuitive user interaction. Previous work has demonstrated AR's effectiveness in surgical training \cite{liu2024exploration} and planning~\cite{fotouhi2020development,song2022happy}, while research in ultrasound visualization has shown that in situ display methods can improve depth perception and task performance compared to conventional screen-based approaches~\cite{bajura1992merging,state1996technologies}. In spine interventions, optical see-through (OST) head-mounted display (HMD)-based AR has been investigated for both needle guidance and surgical navigation. Studies have shown that AR can accelerate and maintain accuracy in facet joint injections~\cite{agten2018augmented,song2026comparative} and improve first-pass success in lumbar puncture when combined with ultrasound visualization~\cite{jiang2023wearable, song2026comparative}. In surgical contexts, AR headsets have been used for pedicle screw placement in phantoms and patients, demonstrating comparable accuracy to fluoroscopy~\cite{gibby2019head} and freehand techniques~\cite{ma2025augmented}, while also reducing placement time. Recent trials further suggest AR navigation can approach the accuracy of robotic systems in pedicle screw fixation~\cite{altorfer2025pedicle}. Despite these advances, most existing systems rely on static CT overlays or manual registration and lack integration with robotics or real-time ultrasound, limiting their adaptability during procedures.

\revision{The CBCT-based AR condition evaluated in the present work was also included in the comparative study by Song~\emph{et al.}~\cite{song2026comparative}, which contrasted CBCT-based AR with an ultrasound shape-completion AR workflow. Accordingly, the participant and trial data from the CBCT-AR condition are shared between the two studies and are not independent new observations. The distinct objective of the present analysis is to evaluate the effect of visualization modality through a within-subject comparison between the CBCT-based AR interface and conventional screen-based guidance.}

\par
Building on these foundations, we present an AR-guided, multi-modality robotic system that combines US and CBCT for spine injections planning and guidance. The system integrates 3D visualization from CBCT, real-time feedback from the robotic US, and AR-based display to enhance anatomical understanding during navigation and needle trajectory planning (Fig.~\ref{fig:teaser}). This design preserves the practicality of RUS while addressing incomplete anatomical context and off-patient visualization.
The key contributions of this work include: (1) development of an AR interface that provides in situ visualization of co-registered CBCT and ultrasound for better navigation during spine interventions, and (2) experimental validation through a within-subject user study, comparing AR to conventional screen-based visualization on representative tasks, evaluated in terms of efficiency, placement accuracy, cognitive workload, usability, trust, and spatial understanding. \revision{The contribution of the evaluation is the paired visualization comparison; the CBCT-AR condition itself shares participant and trial data with Song~\emph{et al.}~\cite{song2026comparative}.}

\section{Methods}\label{sec:Methods}
\begin{figure*}[t]
    \centering
    \includegraphics[width=\textwidth]{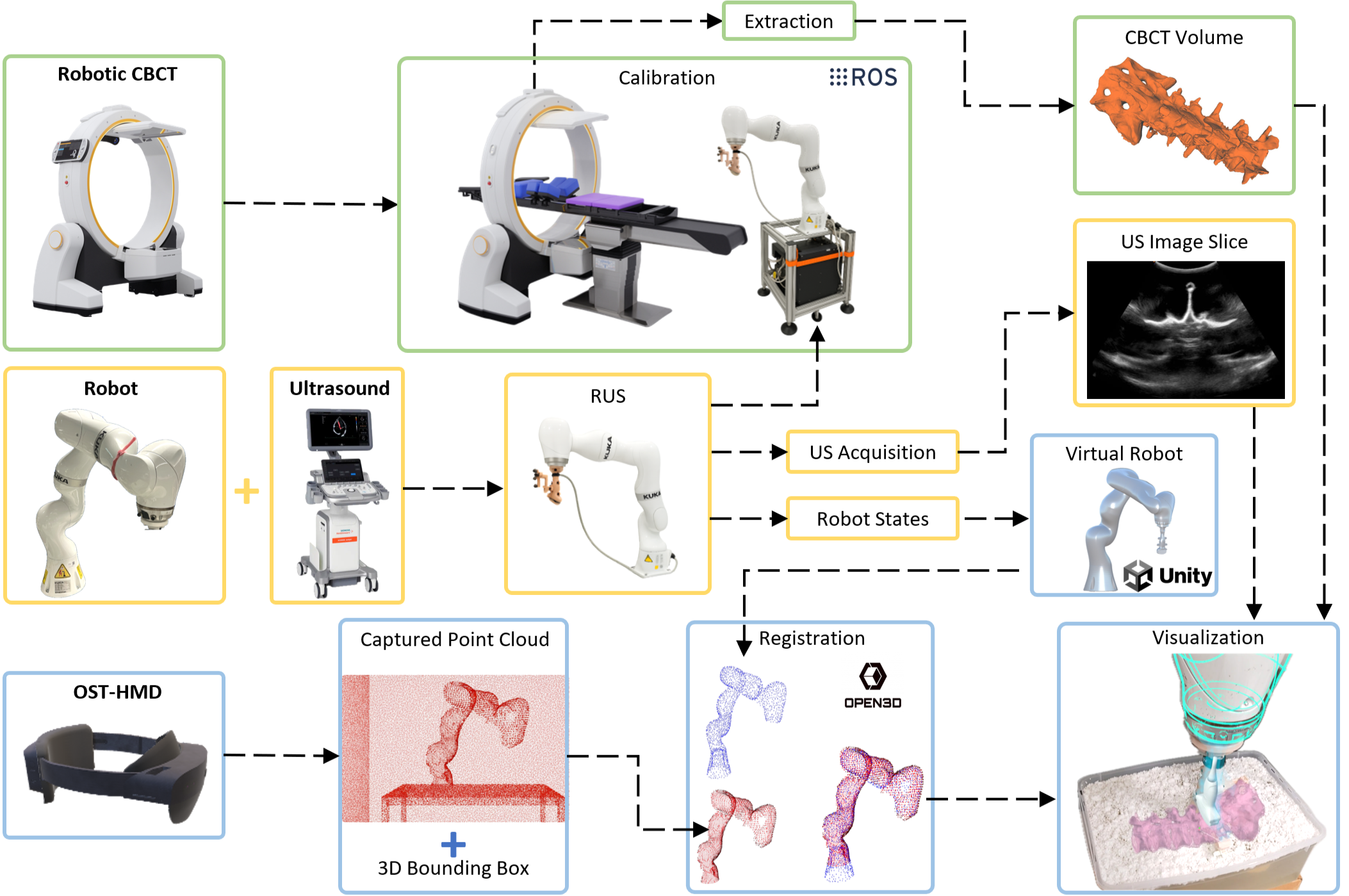}
    \caption{\textbf{System Overview.} The pipeline integrates robotic CBCT, robotic ultrasound, and AR into a unified framework, enabling in situ AR visualization of spine anatomy and needle guidance. {(\color{green} $\boldsymbol{\bullet}$}) represents CBCT volume acquisition, {(\color{yellow} $\boldsymbol{\bullet}$}) depicts ultrasound and robot state acquisition, and {(\color{cyan} $\boldsymbol{\bullet}$}) indicates the visualization process.}
    % Dashed arrows indicate data flow over wireless TCP/IP connections, while solid lines represent internal processing or direct cable connections. }
    \label{fig:overview}
\end{figure*}

The system integrates pre-acquired CBCT volume and robotic ultrasound acquisition with real-time AR visualization to enhance spatial understanding of spinal anatomy. Fig.~\ref{fig:overview} illustrates the overall system architecture and data flow between components.

\subsection{System Pre-Calibration}\label{sec:calibration}
\begin{figure}
    \centering
    \includegraphics[width=0.95\columnwidth]{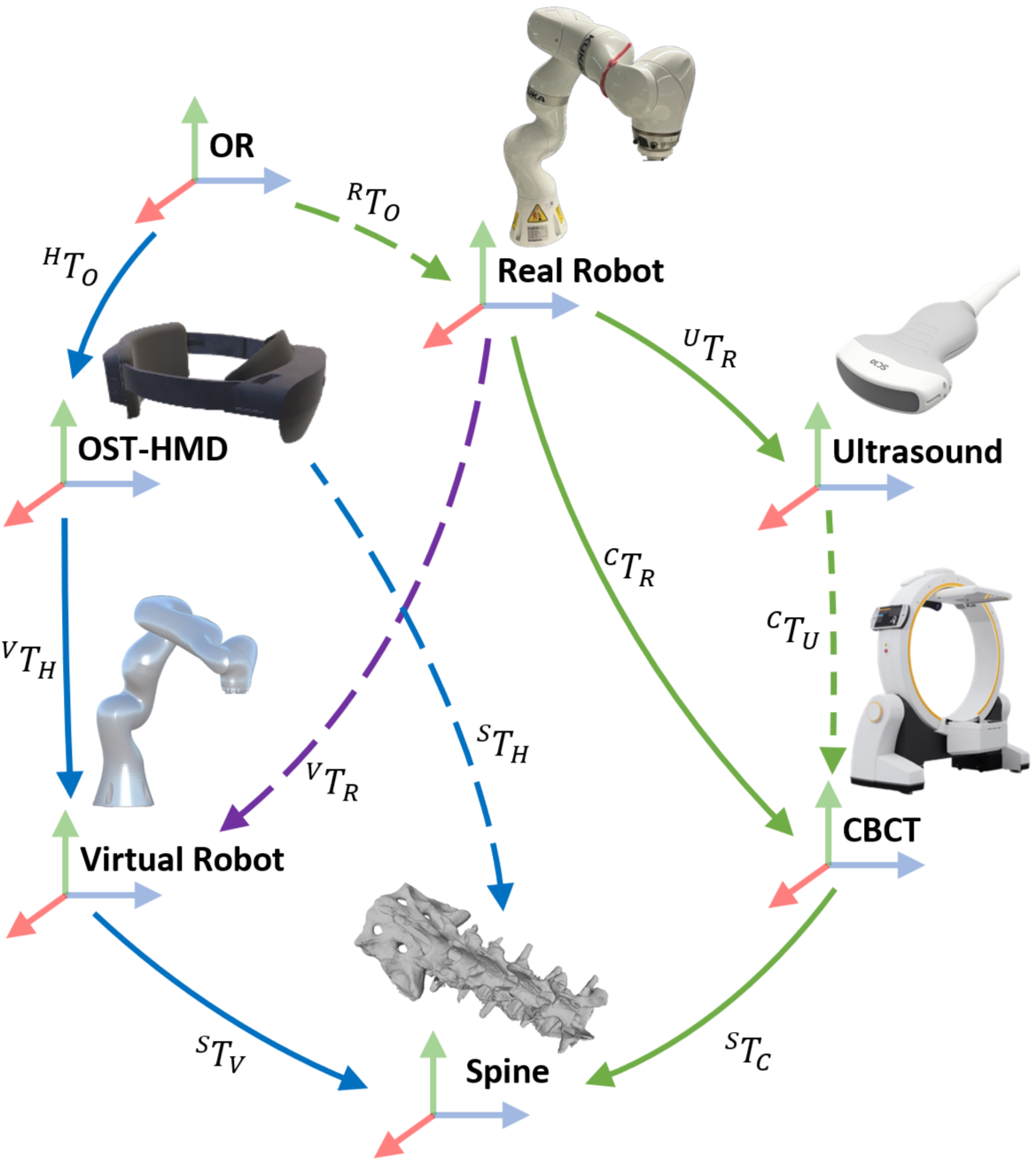}
    \caption{\textbf{Transformation Chain for System Registration.} Note that the transformations shown with solid arrows are acquired, while the transformations with dashed arrows are derived. {(\color{cyan} $\boldsymbol{\bullet}$}) represents transformations in the virtual domain, {(\color{green} $\boldsymbol{\bullet}$}) corresponds to transformations in the real world, and {(\color{violet} $\boldsymbol{\bullet}$}) denotes the registration between virtual and real robot. } 
    \label{fig:chain}
\end{figure}

The purpose of this pre-calibration is to localize the ultrasound image within the CBCT volume using the robot arm, such that after a one-time calibration, the ultrasound image, CBCT slice, and their corresponding spatial transformations can be obtained synchronously in real time. \revision{The coordinate frames are: CBCT image ($C$), robot base ($R$), optical camera ($O$), probe marker ($M$), robot end-effector ($E$), ultrasound image ($U$), virtual robot ($V$), HMD ($H$), and spine model ($S$).}
Following the approach by Li~\emph{et al.} \cite{li2025robotic}, a calibration process between robotic CBCT and the RUS is performed as follows. An optical tracking camera mounted on the CBCT machine tracks a marker rigidly attached to the ultrasound probe, yielding the transformation between the optical camera and the marker $^{O}\mathbf T_{M}$. In parallel, the transformation between the robot end-effector and the robot base $^{E}\mathbf T_{R}$ is directly obtained from the robot control framework~\cite{hennersperger2017towards}. By collecting multiple pairs of corresponding $^{O}\mathbf T_{M}$ and $^{E}\mathbf T_{R}$, a hand–eye calibration problem can be formulated. Specifically, the relationship between consecutive poses can be written as:

\begin{equation} 
\begin{split}
^{E}\mathbf T^{(i)}_{R} \, ^{R}\mathbf T_{O} \, ^{O}\mathbf T^{(i)}_{M} & = ^{E}\mathbf T^{(i+1)}_{R} \, ^{R}\mathbf T_{O} \, ^{O}\mathbf T^{(i+1)}_{M} \\
(^{E}\mathbf T^{(i+1)}_{R})^{-1} \, ^{E}\mathbf T^{(i)}_{R} \, ^{R}\mathbf T_{O} & = ^{R}\mathbf T_{O} \, ^{O}\mathbf T^{(i+1)}_{M} \, (^{O}\mathbf T^{(i)}_{M})^{-1} \\
% A\mathbf X & = \mathbf XB
\end{split}
\label{eq1}
\end{equation}

where $^{R}\mathbf{T}_{O}$ denotes the transformation between the robot base and the optical tracking camera. This formulation corresponds to the classical hand–eye calibration problem ($\mathbf A \mathbf X = \mathbf X\mathbf B$), which can be solved using established optimization methods. Once $^{R}\mathbf T_{O}$ is obtained, the transformation $^{C}\mathbf T_{R}$, as illustrated in Fig.~\ref{fig:chain}, can then be computed as:

\begin{equation} 
^{C}\mathbf T_{R} = ^{C}\mathbf T_{O}(^{R}\mathbf T_{O})^{-1},
\end{equation}

where $^{C}\mathbf T_{O}$ denotes an intrinsic parameter of the robotic CBCT machine~\cite{karius2024first}, while the ultrasound calibration matrix $^{U}\mathbf T_{R}$ is obtained seperately following the approach described by Jiang \emph{et al.}~\cite{jiang2021autonomous}. The transformation between the ultrasound and the CBCT volume, $^{C}\mathbf T_{U}$, is then given by:

\begin{equation} 
^{C}\mathbf T_{U} = ^{C}\mathbf T_{R}(^{U}\mathbf T_{R})^{-1}.
\end{equation}

\revision{The transformation between the patient (spine) coordinate system and the CBCT device, denoted as $^{S}\mathbf T_{C} \in SE(3)$, is determined during the scan and can be computed from the DICOM metadata, which defines the CBCT volume. To enable efficient visualization of the spine model in the HMD, the mesh model was extracted from the DICOM volume using 3D Slicer~\cite{fedorov20123d}, with the anchor point defined at the spine coordinate origin as specified in the DICOM header.}

\subsection{Virtual-to-Real Registration} 
Accurate spatial registration between virtual and physical elements is fundamental to the system's effectiveness. Prior work has explored AR-to-robot registration using marker-based methods, highlighting the importance of registration parameters for achieving reliable alignment accuracy~\cite{mielke2025enhancing}. Other approaches have leveraged point cloud–based registration. While effective, such methods often suffer from long runtimes and potential errors when similar objects are present in the scene~\cite{ostanin2020human}. To mitigate this, some systems require users to provide an initial manual alignment~\cite{kerkhof2025depth}.
In contrast, our method builds on these insights but introduces a streamlined registration pipeline that enables fast and robust registration even in cluttered environments.

The registration process begins with the user's definition of a 3D bounding box around the physical robot to filter environmental noise from depth acquisitions. Users capture the robot from multiple viewpoints while the system accumulates depth frames, creating a point cloud representation of the physical robot $P_{\text{real}} = \{\mathbf{p}_i \in \mathbb{R}^3 \mid i = 1, \dots, N\}$.
A corresponding virtual point cloud $P_{\text{virtual}} = \{\mathbf{q}_i \in \mathbb{R}^3 \mid i = 1, \dots, N\}$ is sampled from the surface of the virtual robot, with joint configurations synchronized to match the physical robot's current state. Consequently, the registration process can be performed at any time without the need to position the robot into specific, predefined poses. The registration transformation $^{V}\mathbf T_{R}^{\star}$ is computed in a two-step process: first, a global registration is performed using maximal cliques for 3D registration~\cite{zhang20233d, yang2024mac}, and then the alignment is refined with a point-to-point iterative closest point (ICP) optimization.

\begin{equation}
    ^{V}\mathbf T_{R}^{\star}
= \arg\min_{\,^{V}\mathbf T_{R}\in SE(3)}
\sum_{i}\big\|
\,^{V}\mathbf T_{R}\,\mathbf p_i - \mathbf q_{\pi(i)}
\big\|^2,
\end{equation}

where $\pi(i)$ denotes the correspondence mapping between real and virtual points.

Following the calibration in Sect.~\ref{sec:calibration}, the CBCT volume is co-registered with the real robot coordinate system through the known transforms $^{C}\mathbf{T}_R$. The reconstructed spine mesh is extracted from the CBCT, yielding a fixed transform $^{S}\mathbf{T}_C$. By chaining the registration and transforms, the reconstructed spine can finally be expressed in the OST-HMD coordinate system as
\begin{equation}
^{S}\mathbf{T}_H
= \; ^{S}\mathbf{T}_C \; ^{C}\mathbf{T}_R \; (^{V}\mathbf{T}_R)^{-1} \; ^{V}\mathbf{T}_H
\end{equation}

\subsection{AR Guidance Interface}

The AR interface provides in situ visualization of the registered spine model together with an interactive needle guidance overlay. The needle guide axis is modeled as a ray in the robot coordinate system. 
Let the needle guide origin in the robot base frame be $\mathbf{o}_R \in \mathbb{R}^3$ and its unit direction vector be $\mathbf{d}_R \in \mathbb{R}^3$. The virtual needle trajectory is then parameterized as

\begin{equation}
\mathbf{r}(t) = \mathbf{o}_R + t\,\mathbf{d}_R, \quad t \geq 0.
\end{equation}

The CBCT-derived 3D spine model is represented as a triangular mesh. A ray–triangle intersection test is performed between $\mathbf{r}(t)$ and the mesh using the Möller–Trumbore algorithm. The smallest positive intersection parameter $t^\star$ yields the predicted contact point

\begin{equation}
\mathbf{p}_{\text{hit}} = \mathbf{o}_R + t^\star \mathbf{d}_R.
\end{equation}

\begin{figure}
    \centering
    \includegraphics[width=0.6\columnwidth]{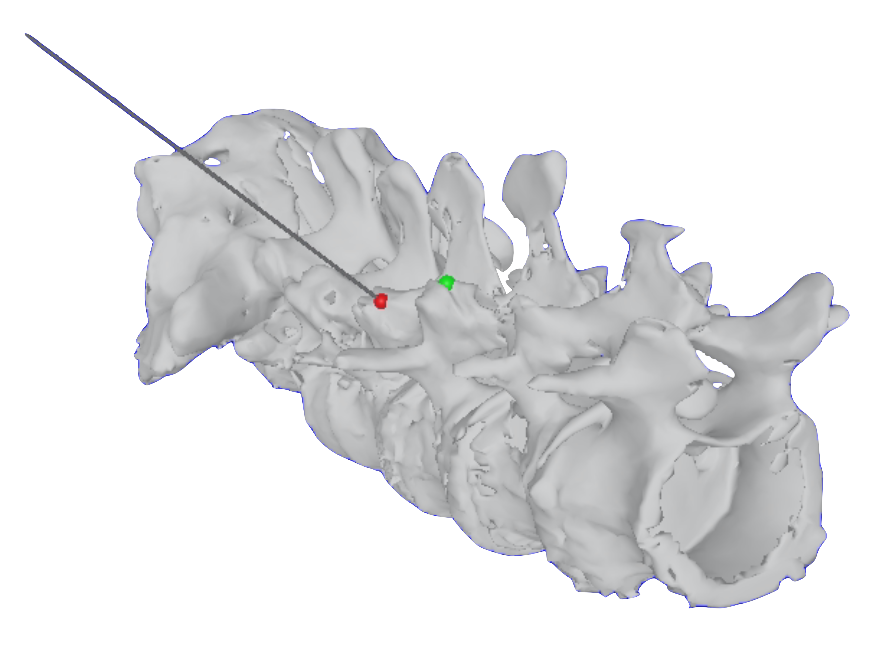}
    \caption{\textbf{AR Guidance Visualization.} Raycasting from the needle guide onto the CT-derived spine model yields the predicted intersection point (marked in red) between the needle and the spine model.} 
    \label{fig:unity}
\end{figure}

This hit point is rendered in situ as a small red sphere overlaid on the spine model, providing intuitive feedback of the expected anatomical contact point (Fig.~\ref{fig:unity}). If no intersection is detected, no hit point is displayed, ensuring that users are only presented with anatomically valid trajectories.
This combination of real-time raycasting and in situ visualization allows users to quickly infer where the needle will intersect the anatomy, while live ultrasound remains available to verify tip placement during insertion.

\revision{The CBCT-AR observations analyzed in the present study were previously reported as one condition in Song~\emph{et al.}~\cite{song2026comparative}. The present manuscript uses these observations together with the conventional screen-based condition to address the paired AR-versus-screen visualization comparison.}
\section{Experiments}\label{Sec:UserStudy}

To evaluate the feasibility of the proposed AR-guided robotic ultrasound system, we conducted a user study that was designed to demonstrate the system’s applicability to two representative spine procedures: facet joint \revision{FJ} injection and lumbar puncture \revision{LP}. 
% Both tasks were implemented on the same experimental platform and followed a within-subject design, where each participant performed the procedures under both AR-guided and conventional screen-based visualization conditions.

\subsection{Setup}
Our experimental setup consists of a KUKA LBR iiwa 14 R820 robotic arm and a Siemens ACUSON Juniper ultrasound machine. The Siemens 5C1 convex ultrasound probe is rigidly attached to the robot’s end-effector using a 3D-printed probe holder. The probe holder also incorporated a detachable needle guide to simulate injection planning. A detachable tracking marker was mounted on the probe holder during pre-calibration and subsequently removed for the user study to provide an unobstructed view. The robotic arm is operated using a handheld game controller, which allows intuitive manipulation of the probe’s position and orientation during scanning and needle placement tasks.
For anatomical reference, we employed a lumbar spine phantom embedded in a 35 cm $\times$ 25 cm $\times$ 24 cm container, and there is a layer of gel at the bottom to fix the spine. The phantom was positioned such that the superior portion of the lumbar spine was submerged in water, providing acoustic coupling for ultrasound imaging. To prevent participants from directly viewing the phantom anatomy, small styrofoam pellets were placed on the water surface, ensuring that task performance relied solely on the imaging and AR guidance. %This configuration enabled consistent image quality while preserving realistic anatomical landmarks relevant to our chosen spine procedures.
AR visualization was implemented in Unity and delivered through a Microsoft HoloLens 2 HMD. Communication between the robotic system and the HMD was handled via TCP/IP. To enable quantitative evaluation of needle placement, \revision{the needle tip was manually localized after task completion by an experimenter, who selected its position with a mouse in the final frozen frame of the two-dimensional ultrasound image, which displays the needle tips on the monitor. The annotator did not know the predefined target coordinate during localization, and the same ultrasound visualization and windowing were used for every trial. The selected image position was transformed from the calibrated ultrasound image coordinate system into the 3D CBCT coordinate system using the pre-calibrated US--robot and robot--CBCT transformations. Placement error was defined as the three-dimensional Euclidean distance between the resulting needle-tip position and the predefined target point in the CBCT volume. A single manual annotation was used for each trial; annotation repeatability and inter-rater variability were not evaluated and therefore constitute an additional unquantified source of measurement uncertainty.}

\begin{figure}[b]
    \centering
    \includegraphics[width=0.485\textwidth]{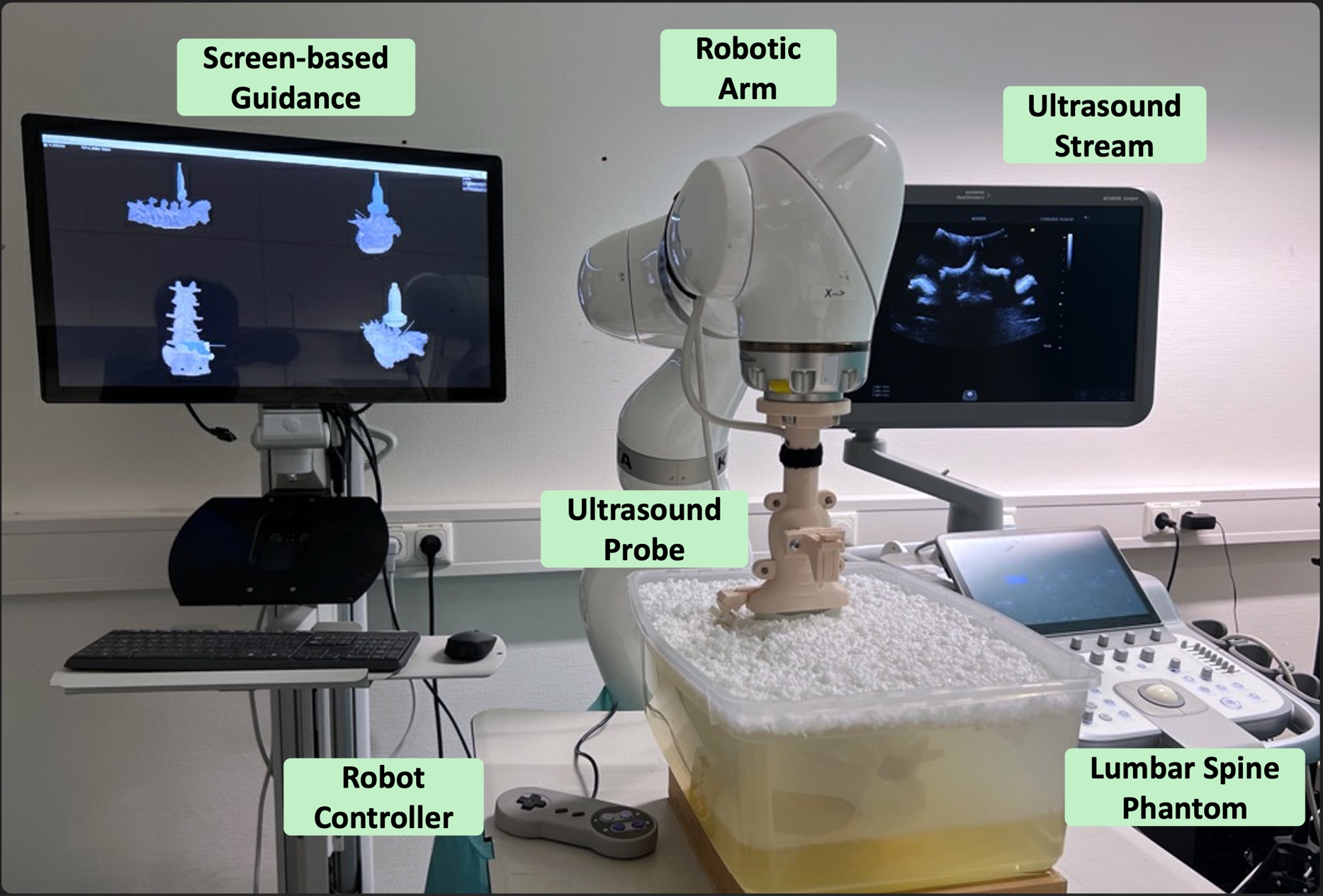}
        \caption{\textbf{Study Setup.} Experimental configuration with robotic arm, ultrasound probe, lumbar spine phantom, and screen-based guidance interface.}
    \label{fig:setup}
\end{figure}

% we co-calibrated the robotic ultrasound system with the Brainlab LoopX mobile cone-beam CT (CBCT) scanner following the method proposed by Li et al.~\cite{li2025robotic}. As a result, needle placement could be evaluated directly from ultrasound acquisitions by transforming the localized needle tip into the 3D CT coordinate system.
% The experimental station was arranged on a clinical workbench, with participants standing at a comfortable working distance from the phantom. The ultrasound console was placed adjacent to the phantom to allow conventional 2D monitor-based viewing during the baseline condition. The AR and screen-based interfaces were alternated according to the study design.

\subsection{Participants}
A total of 16 subjects (6 female, 10 male), aged 22-43 (M = 28.75, SD = 4.49) years, participated in the study. All participants reported normal or corrected-to-normal vision and no known musculoskeletal impairments that might interfere with task execution. Participants rated their task-relevant experience on a 5-point Likert scale, where 1 indicated no familiarity and 5 indicated high familiarity. They reported moderate familiarity with AR technologies ($3.38 \pm 1.27$), ultrasound experience ($3.44 \pm 1.00$), and robotic systems ($3.56 \pm 1.06$). \revision{None of the participants were practicing clinicians. The cohort was therefore intended to support an initial technical and usability feasibility assessment rather than an evaluation of clinical performance.}

\subsection{Tasks}
The study employed a within-subject design in which each participant performed needle insertion tasks under two visualization conditions: conventional screen-based display and AR-based in situ visualization. The order of conditions was counterbalanced across participants to mitigate learning effects.

Within each condition, participants carried out two insertion trials: one FJ injection (e.g., left L2–L3) and one LP (e.g., L3–L4 interspinous midline). For each trial, the specific target joint or interspinous level was randomly assigned\revision{; each target was only encountered once during the study}. Participants used the robot controller to move the ultrasound probe and the needle guide integrated into the probe holder to align the trajectory toward the designated target\revision{; participants were familiarized with the control methods but did not perform practice trials}.

During all tasks, participants had access to the CBCT-derived 3D spine model together with the live ultrasound feed, enabling them to integrate volumetric anatomical context with real-time imaging feedback when performing the insertion.

\subsection{Procedure and Measurements}

At the beginning of each session, the experimenter introduced the study background and explained the overall procedure. Participants then provided written informed consent and completed a demographic questionnaire. A short tutorial phase followed, during which participants practiced under both visualization conditions until they felt comfortable with the setup.

Each participant performed two insertion trials per condition. For each trial, execution time was recorded as well as the needle placement accuracy. \revision{Execution time started immediately before the participant initiated any movement of the robotic system and ended when the participant verbally declared that the task was complete.} After completing each condition, participants filled out a series of subjective questionnaires: the System Usability Scale (SUS)\cite{brooke1996sus} to assess usability, the NASA Task Load Index (NASA-TLX)\cite{hart2006nasa} to evaluate perceived workload, and the Single Ease Question (SEQ)\cite{sauro2009comparison} to capture task difficulty on a 7-point Likert scale (1 = very difficult, 7 = very easy). To further characterize user experience, we also included a short Trust in Automation scale adapted from Jian et al.~\cite{jian2000foundations}, and a spatial understanding questionnaire adapted from the Igroup Presence Questionnaire (IPQ)~\cite{schubert2001experience}. The full experimental session lasted approximately 40 minutes per participant.

\section{Results}\label{Sec:Results}

\subsection{System Accuracy Evaluation}

Point cloud registration between the real and virtual robot achieved a mean residual (RMSE) of 9.1 mm. The co-calibration between the robotic ultrasound system and the CBCT scanner resulted in a target registration error of $1.72\pm0.62$ mm, evaluated on fiducial markers embedded in the phantom. Ultrasound probe calibration with respect to the robot end-effector yielded a localization error of $0.65\pm0.2$ mm, consistent with values reported in previous robotic ultrasound systems. \revision{The residual surface-alignment RMSE between the acquired physical-robot point cloud and the virtual robot model is not directly equivalent to a target registration error at the needle tip and should not be numerically compared with the fiducial-based US--CBCT TRE or probe-calibration error. The point-cloud transform affects the placement of virtual anatomy and guidance elements in the HMD coordinate frame; however, its contribution to target-space error was not isolated in the present study.}

\subsection{Study Statistics}
Statistical analyses were conducted to assess the significance of differences between AR and screen-based conditions. Data normality was first evaluated using the Shapiro–Wilk test. Measures that met normality assumptions were compared using paired t-tests, while non-normal measures were analyzed with the Wilcoxon signed-rank test. \revision{The significance level was set to $\alpha=0.05$. For paired $t$-tests, $t$ denotes the test statistic, and the corresponding degrees of freedom are reported with each result. For Wilcoxon tests, $W$ denotes \textbf{CONVENTION USED BY THE ORIGINAL SOFTWARE}; $p$-values were computed using \textbf{EXACT METHOD AND ZERO/TIE HANDLING}. All tests were \textbf{SIDEDNESS} and analyses were performed using \textbf{SOFTWARE/VERSION}.}

\revision{Normality was assessed for the paired condition differences using the Shapiro--Wilk test. Normally distributed paired differences were analyzed using paired $t$-tests, whereas non-normal paired differences were analyzed using Wilcoxon signed-rank tests. The significance level was set to $\alpha=0.05$. For paired $t$-tests, $t$ denotes the test statistic and the corresponding degrees of freedom are reported with each result. For Wilcoxon tests, $W$ denotes \textbf{TODO: DEFINE THE CONVENTION USED BY THE ORIGINAL SOFTWARE}; $p$-values were computed using \textbf{TODO: ORIGINAL EXACT/ASYMPTOTIC METHOD AND ZERO/TIE HANDLING}. All tests were \textbf{TODO: ORIGINAL SIDEDNESS} and analyses were performed using \textbf{TODO: SOFTWARE/VERSION}.}

\subsubsection{User Performance}

% \begin{figure}[!t]
%     \centering
%     \includegraphics[width=\columnwidth]{figures/performance.png}
%     \caption{\textbf{Execution Performance.} \textbf{AR} yielded faster performance for both tasks and significantly lower error for LP.}
%     \label{fig:performance}
% \end{figure}

\begin{figure}[!t]
    \centering
    \includegraphics[width=\columnwidth]{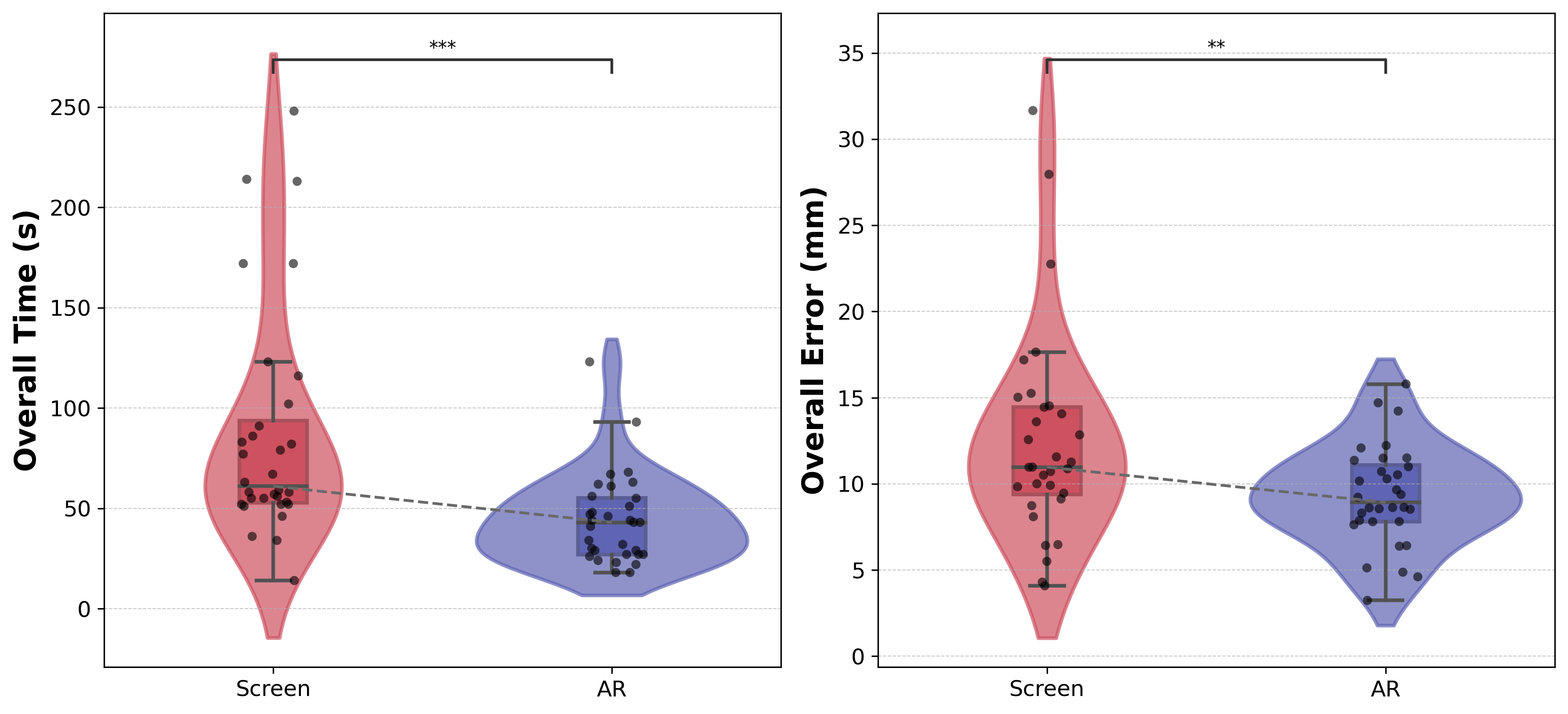}
    \caption{\textbf{Overall Across-Task Performance.} Significance: $ \star p<.05$, $\star\star p<.01$, $\star\star\star p<.001$.}
    \label{fig:performance_overall}
\end{figure}

% Participants completed both procedures faster with \textbf{AR} than with Screen. Results are summarized in Table~\ref{tab:objective}. % and visualized in Fig.~\ref{fig:performance}.
% For execution time, participants were faster with \textbf{AR} than \textbf{Screen} for both tasks:
% facet joint (FJ) showed a 54\% reduction (\(W=13.0,\, p=0.003\)), and lumbar puncture (LP) a 42\% reduction (\(W=11.5,\, p=0.002\)). Across tasks, the combined execution time was reduced by 49\% (\(W=5.0,\, p=0.0003\)).

% For placement accuracy, AR yielded a 18\% error reduction for FJ (\(t=1.26,\, p=0.228\)) and a \textbf{32\%} reduction for LP (\(t=2.74,\, p=0.015\)). The combined accuracy improved by 25\% overall (\(t=3.29,\, p=0.0050\)).

\revision{Participants completed both tasks significantly faster with AR than with screen-based guidance. In the existing overall analysis across both tasks, placement error was also significantly lower with AR. In the task-specific analyses, however, the placement-error reduction was statistically significant for LP but not for FJ targeting.} Overall execution time decreased by 49\% (\(W=5.0,\,p=0.0003\)) and placement error decreased by 25\% (\(t=3.29,\,p=0.0050\)) across tasks; see Fig.~\ref{fig:performance_overall}. 
Breaking this down by task, in AR condition, \revision{(FJ)} time decreased by 54\% (\(W=13.0,\,p=0.003\)) and \revision{(LP)} time by 42\% (\(W=11.5,\,p=0.002\)). 
For placement accuracy, AR yielded a 18\% error reduction for FJ (\(t=1.26,\, p=0.228\)) and a 32\% reduction for LP (\(t=2.74,\, p=0.015\)).
Full per-task statistics are reported in Table~\ref{tab:objective}.

\begin{table}[b]
\centering
\caption{\textbf{Task Specific Results (mean$\pm$SD).} Significance is indicated in bold.}
\label{tab:objective}
\footnotesize
\setlength{\tabcolsep}{9pt}
\begin{tabular}{lccc}
\toprule
\textbf{Measure} & \textbf{Screen} & \textbf{AR} & \textbf{$p$ value} \\
\midrule
FJ time (s)            & 95.25$\pm$54.51 & 43.50$\pm$19.17 & \textbf{0.003} \\
FJ error (mm)            & 11.76$\pm$6.37  & 9.69$\pm$2.93   & 0.228 \\
LP time (s)            & 78.25$\pm$56.46 & 45.31$\pm$24.90 & \textbf{0.002} \\
LP error (mm)            & 13.12$\pm$5.43  & 8.89$\pm$2.71   & \textbf{0.015} \\
% \midrule
% Combined time (s)      & 86.75$\pm$56.14 & 44.41$\pm$22.24 & 0.0003** \\
% Combined error (mm)      & 12.44$\pm$5.96  & 9.29$\pm$2.85   & 0.0050** \\
\bottomrule
\end{tabular}
\end{table}

\subsubsection{Subjective Ratings}
\begin{figure*}[!t]
    \centering
    \includegraphics[width=\textwidth]{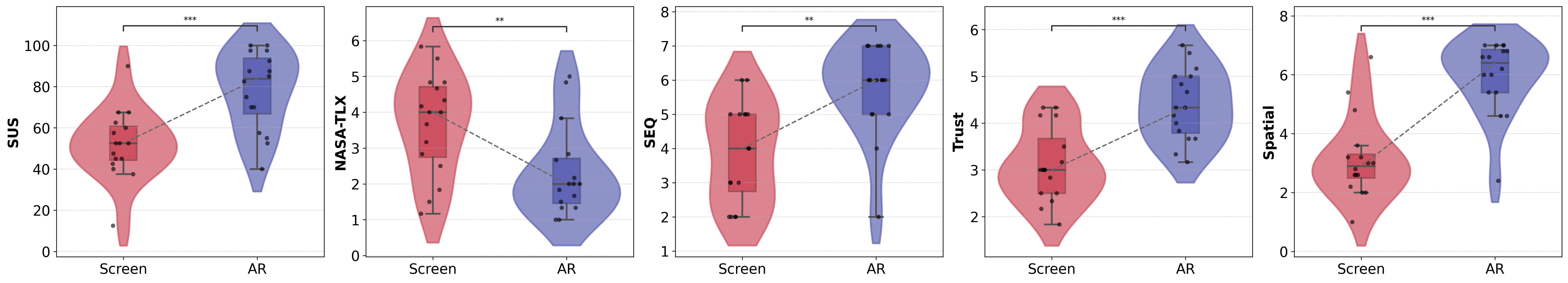}
    \caption{\textbf{Subjective Ratings.} AR significantly increased usability, ease of use, trust, and spatial understanding, and reduced cognitive workload.}
    \label{fig:subjective}
\end{figure*}

Subjective ratings (Fig.~\ref{fig:subjective}) favored AR across all measures: usability was higher by 50\% (\(t=-4.33,\, p=0.00060\)); workload was lower by 37\% (\(t=3.63,\, p=0.0025\)); perceived ease increased by 48\% (\(W=5.00,\, p=0.0010\)); trust rose by 41\% (\(t=-4.25,\, p=0.00070\)); and spatial understanding showed the largest gain, improving by 89\% (\(t=-5.70,\, p=0.0004\)).

\section{Discussion}\label{Sec:Discussion}

Overall, participants performed faster and more accurately with AR than with the conventional screen, and their subjective ratings moved in the same direction—higher usability, trust, and spatial understanding with lower perceived workload and difficulty. These findings are consistent with prior phantom studies showing that AR in situ US reduces time and needle passes compared to standard ultrasound guidance~\cite{farshad2020ultrasound,von2022ultrarsound}.
A plausible explanation is that the in situ CBCT model, rendered in the same coordinate frame as the robot and ultrasound, externalizes the 3D spatial reasoning that users otherwise have to map mentally. Live ultrasound then serves as a local, real-time check on the needle tip while the CBCT provides a stable global 3D map, which likely explains the large time savings. 
% The accuracy gains were more pronounced for lumbar puncture than for facet joints, consistent with ultrasound offering clearer landmarks for interspinous access, whereas facet targets are smaller, often shadowed, and approached at oblique angles—conditions under which AR still helps with orientation but cannot fully compensate for imaging limitations.

\revision{The task-specific analyses yielded different significance outcomes: the placement-error reduction was statistically significant for LP but not for FJ targeting. This difference in significance does not by itself demonstrate that the effect of AR was statistically different between the two tasks. One possible explanation for the observed pattern is that the midline interspinous trajectory benefits more directly from the displayed trajectory and intersection cue, while the oblique facet-joint approach and local acoustic shadowing may limit the utility of the overlay. This interpretation remains hypothetical because the study was not designed to isolate the influence of target geometry.}

\revision{The} observed \revision{FJ} targeting errors are within the anatomical dimensions of the lumbar facets, which typically measure 14–19 mm in width and 17–20 mm in height~\cite{gorniak2015lower,esteban2018robotic}. This indicates that the placement deviations reported in our study remain within a clinically acceptable range for \revision{FJ} interventions. Anatomical studies also indicate that interspinous distances (7–10 mm) and spinous process dimensions at L3–L4 are small enough that localization errors larger than this can compromise safety or efficacy~\cite{neumann1999determination, feng2020morphological}. \revision{These dimensions provide geometric context for the observed errors, but do not establish a reliable safety margin. The observed AR placement errors for both procedures do not establish suitability for clinical use, particularly because nearby critical structures, patient-specific anatomy, needle bending, tissue deformation, respiration, patient movement, or posture changes occurring after CBCT acquisition were not represented in the rigid phantom experiment. Any resulting change in the relationship between the anatomy and the pre-acquired CBCT model would introduce registration drift.}

\revision{An isolated system-level end-to-end TRE of the AR guidance and display chain, from the physical/CBCT anatomy through registration and tracking to HMD rendering, was not measured. The user-study placement error is a different endpoint: it combines participant execution with technical system inaccuracies and the separate needle-tip outcome-measurement chain, including manual ultrasound localization and transformation into CBCT coordinates. Placement error therefore should not be interpreted as a direct measurement of full-chain AR TRE. Future technical validation should quantify guidance/display TRE independently using rigid tracked targets distributed throughout the relevant working volume and repeated robot-to-HMD registrations.}

\revision{Vergence--accommodation conflict, imperfect virtual--real occlusion, and other optical see-through depth-perception effects were not formally assessed. Although no corresponding questionnaire data were collected, these display-related factors may influence manual alignment precision and should be evaluated prospectively.}

\revision{The apparent relative jitter between the physical and virtual robot in the supplementary video was introduced by the mixed-reality recording and compositing process. It was not visible to participants in the optical see-through AR view during task execution. This recording artifact is not representative of the visualization experienced during the study. No separate quantitative evaluation of HMD tracking stability was performed.}

\revision{Because the participants were not practicing clinicians, the observed improvements cannot be assumed to generalize to interventional radiologists, anesthesiologists, or spine surgeons, whose anatomical knowledge and familiarity with conventional guidance may change both baseline performance and the benefit obtained from AR. Structured evaluation with clinical experts and subsequent cadaveric or clinical validation are required.}

\revision{Task completion was determined by each participant's verbal declaration rather than an independent anatomical or image-based stopping rule. Although this criterion was applied consistently across conditions, individual confidence and decision thresholds may have contributed to variability in the measured execution times.}

\revision{Direct numerical comparison with previous systems must be interpreted cautiously because the reported endpoints differ. Jiang~\emph{et al.}~\cite{jiang2023wearable} reported below-3-mm overall navigation accuracy in a dedicated technical evaluation, whereas our placement-error measure combines system inaccuracies with participant execution and manual tip-localization error. Agten~\emph{et al.}~\cite{agten2018augmented} reported 39 of 40 AR-guided facet injections as perfect or acceptable, with one placement missing the joint by 2~mm, rather than reporting a directly equivalent mean user-targeting error. These findings demonstrate improved performance relative to the screen-based baseline under the controlled conditions of this phantom study.} 

\revision{Further studies, however, need to accommodate for the deformation effects stated above, in order to ensure generalizability to complex scenarios \cite{li2026robotic}. A clinical translation would therefore likely require mechanisms such as dynamic anatomical tracking or periodic US-to-CBCT re-registration.}

A promising future research direction is to compare our workflow with emerging ultrasound-only approaches for spine shape completion \cite{gafencu2024shape, gafencu2025shape}. Such methods could eliminate radiation exposure and streamline the procedure. A direct comparison would clarify the trade-offs between CBCT-derived models and purely ultrasound-based reconstructions.

\section{Conclusion}\label{Sec:Conclusion}

We introduced an AR-guided robotic ultrasound workflow for enhanced visualization and guidance during needle placement in spine interventions. 
Our proposed method integrates together the robotic ultrasound system, the CBCT-derived 3D model, and the HMD in a single registered coordinate frame. In a within-subject phantom study covering two use cases—facet joint injection and lumbar puncture—participants completed insertions under AR and conventional screen displays. AR yields faster execution and lower placement error across tasks, with consistent per-task time savings and a clear accuracy benefit for lumbar puncture. Subjective ratings aligned with these outcomes: AR improved usability, trust, and spatial understanding while reducing perceived workload and task difficulty.
These findings support the feasibility of in-situ AR guidance for general spine interventions, where a 3D model provides global context and live ultrasound supplies local, real-time verification. 
% Future work will validate the workflow with clinicians on cadavers and patients, incorporate online needle-tip tracking and continuous calibration quality checks, study ergonomics and longer procedures, and extend the approach to additional spine targets such as epidural access and pedicle instrumentation.

\bibliography{wileyNJD-Chicago}

@article{hennersperger2017towards,
  title={{T}owards {MRI}-based autonomous robotic {US} acquisitions: a first feasibility study},
  author={Hennersperger, Christoph and Fuerst, Bernhard and Virga, Salvatore and Zettinig, Oliver and Frisch, Benjamin and Neff, Thomas and Navab, Nassir},
  journal={IEEE transactions on medical imaging},
  volume={36},
  number={2},
  pages={538--548},
  year={2017},
  publisher={IEEE}
}

@article{gafencu2024shape,
  title={{S}hape completion in the dark: completing vertebrae morphology from 3{D} ultrasound},
  author={Gafencu, Miruna-Alexandra and others},
  journal={IJCARS},
  volume={19},
  number={7},
  pages={1339--1347},
  year={2024},
  publisher={Springer}
}

@article{wu2016effectiveness,
  title={{E}ffectiveness of ultrasound-guided versus fluoroscopy or computed tomography scanning guidance in lumbar facet joint injections in adults with facet joint syndrome: a meta-analysis of controlled trials},
  author={Wu, Tao and Zhao, Wei-hua and Dong, Yan and Song, Hai-xin and Li, Jian-hua},
  journal={Archives of physical medicine and rehabilitation},
  volume={97},
  number={9},
  pages={1558--1563},
  year={2016},
  publisher={Elsevier}
}

@article{evansa2015ultrasound,
  title={{U}ltrasound versus fluoroscopic-guided epidural steroid injections in patients with degenerative spinal diseases: a randomised study},
  author={Evansa, Irina and Logina, Inara and Vanags, Indulis and Borgeat, Alain},
  journal={European Journal of Anaesthesiology| EJA},
  volume={32},
  number={4},
  pages={262--268},
  year={2015},
  publisher={LWW}
}

@inproceedings{zhang20233d,
  title={3D Registration with Maximal Cliques},
  author={Zhang, Xiyu and Yang, Jiaqi and Zhang, Shikun and Zhang, Yanning},
  booktitle={Proceedings of the IEEE/CVF Conference on Computer Vision and Pattern Recognition},
  pages={17745--17754},
  year={2023}
}

@article{yang2024mac,
  title={MAC: Maximal Cliques for 3D Registration},
  author={Yang, Jiaqi and Zhang, Xiyu and Wang, Peng and Guo, Yulan and Sun, Kun and Wu, Qiao and Zhang, Shikun and Zhang, Yanning},
  journal={IEEE Transactions on Pattern Analysis and Machine Intelligence},
  year={2024},
  publisher={IEEE}
}

@article{gafencu2025shape,
  title={Shape Completion and Real-Time Visualization in Robotic Ultrasound Spine Acquisitions},
  author={Gafencu, Miruna-Alexandra and Shaban, Reem and Velikova, Yordanka and Azampour, Mohammad Farid and Navab, Nassir},
  journal={arXiv preprint arXiv:2508.08923},
  year={2025}
}

@article{jian2000foundations,
  title={Foundations for an empirically determined scale of trust in automated systems},
  author={Jian, Jiun-Yin and Bisantz, Ann M and Drury, Colin G},
  journal={International journal of cognitive ergonomics},
  volume={4},
  number={1},
  pages={53--71},
  year={2000},
  publisher={Taylor \& Francis}
}

@article{schubert2001experience,
  title={The experience of presence: Factor analytic insights},
  author={Schubert, Thomas and Friedmann, Frank and Regenbrecht, Holger},
  journal={Presence: Teleoperators \& Virtual Environments},
  volume={10},
  number={3},
  pages={266--281},
  year={2001},
  publisher={MIT Press One Rogers Street, Cambridge, MA 02142-1209, USA journals-info~…}
}

@article{li2025robotic,
  title={Robotic CBCT meets robotic ultrasound},
  author={Li, Feng and Bi, Yuan and Huang, Dianye and Jiang, Zhongliang and Navab, Nassir},
  journal={International Journal of Computer Assisted Radiology and Surgery},
  pages={1--9},
  year={2025},
  publisher={Springer}
}

@inproceedings{sauro2009comparison,
  title={Comparison of three one-question, post-task usability questionnaires},
  author={Sauro, Jeff and Dumas, Joseph S},
  booktitle={Proceedings of the SIGCHI conference on human factors in computing systems},
  pages={1599--1608},
  year={2009}
}

@inproceedings{hart2006nasa,
  title={NASA-task load index (NASA-TLX); 20 years later},
  author={Hart, Sandra G},
  booktitle={Proceedings of the human factors and ergonomics society annual meeting},
  volume={50},
  number={9},
  pages={904--908},
  year={2006},
  organization={Sage publications Sage CA: Los Angeles, CA}
}

@article{brooke1996sus,
  title={SUS: A quick and dirty usability scale},
  author={Brooke, J},
  journal={Usability Evaluation in Industry},
  year={1996}
}

@article{bi2024machine,
  title={Machine learning in robotic ultrasound imaging: Challenges and perspectives},
  author={Bi, Yuan and Jiang, Zhongliang and Duelmer, Felix and Huang, Dianye and Navab, Nassir},
  journal={Annual Review of Control, Robotics, and Autonomous Systems},
  year={2024},
  publisher={Annual Reviews}
}

@article{rasoulian2015ultrasound,
  title={Ultrasound-guided spinal injections: a feasibility study of a guidance system},
  author={Rasoulian, Abtin and Seitel, Alexander and others},
  journal={IJCARS},
  volume={10},
  number={9},
  pages={1417--1425},
  year={2015},
  publisher={Springer}
}

@article{tirindelli2020force,
  title={Force-ultrasound fusion: Bringing spine robotic-us to the next “level”},
  author={Tirindelli, Maria and Victorova, Maria and others},
  journal={IEEE Robotics and Automation Letters},
  volume={5},
  number={4},
  pages={5661--5668},
  year={2020},
  publisher={IEEE}
}

@article{wang2023compliant,
  title={Compliant joint based robotic ultrasound scanning system for imaging human spine},
  author={Wang, Yunjiang and Liu, Tianjian and Hu, Xinben and Yang, Keji and Zhu, Yongjian and Jin, Haoran},
  journal={IEEE Robotics and Automation Letters},
  volume={8},
  number={9},
  pages={5966--5973},
  year={2023},
  publisher={IEEE}
}

@article{esteban2018robotic,
  title={Robotic ultrasound-guided facet joint insertion},
  author={Esteban, Javier and Simson, Walter and Requena Witzig, Sebastian and Rienm{\"u}ller, Anna and Virga, Salvatore and Frisch, Benjamin and Zettinig, Oliver and Sakara, Drazen and Ryang, Yu-Mi and Navab, Nassir and others},
  journal={International journal of computer assisted radiology and surgery},
  volume={13},
  number={6},
  pages={895--904},
  year={2018},
  publisher={Springer}
}

@article{song2022happy,
  title={HAPPY: hip arthroscopy portal placement using augmented reality},
  author={Song, Tianyu and Sommersperger, Michael and Baran, The Anh and Seibold, Matthias and Navab, Nassir},
  journal={Journal of Imaging},
  volume={8},
  number={11},
  pages={302},
  year={2022},
  publisher={MDPI}
}

@article{bajura1992merging,
  title={Merging virtual objects with the real world: Seeing ultrasound imagery within the patient},
  author={Bajura, Michael and Fuchs, Henry and Ohbuchi, Ryutarou},
  journal={ACM SIGGRAPH Computer Graphics},
  volume={26},
  number={2},
  pages={203--210},
  year={1992},
  publisher={ACM New York, NY, USA}
}

@inproceedings{state1996technologies,
  title={Technologies for augmented reality systems: Realizing ultrasound-guided needle biopsies},
  author={State, Andrei and Livingston, Mark A and others},
  booktitle={Proceedings of the 23rd annual conference on computer graphics and interactive techniques},
  pages={439--446},
  year={1996}
}

@article{fotouhi2020development,
  title={Development and pre-clinical analysis of spatiotemporal-aware augmented reality in orthopedic interventions},
  author={Fotouhi, Javad and Mehrfard, Arian and Song, Tianyu and Johnson, Alex and Osgood, Greg and Unberath, Mathias and Armand, Mehran and Navab, Nassir},
  journal={IEEE transactions on medical imaging},
  volume={40},
  number={2},
  pages={765--778},
  year={2020},
  publisher={IEEE}
}

@inproceedings{ostanin2020human,
  title={Human-robot interaction for robotic manipulator programming in Mixed Reality},
  author={Ostanin, Mikhail and Mikhel, Stanislav and Evlampiev, Alexey and Skvortsova, Valeria and Klimchik, Alexandr},
  booktitle={2020 IEEE International Conference on Robotics and Automation (ICRA)},
  pages={2805--2811},
  year={2020},
  organization={IEEE}
}

@inproceedings{mielke2025enhancing,
  title={Enhancing AR-to-Robot Registration Accuracy: A Comparative Study of Marker Detection Algorithms and Registration Parameters},
  author={Mielke, Tonia and Heinrich, Florian and Hansen, Christian},
  booktitle={2025 IEEE International Conference on Robotics and Automation (ICRA)},
  pages={4746--4752},
  year={2025},
  organization={IEEE}
}

@article{kerkhof2025depth,
  title={Depth-based registration of 3D preoperative models to intraoperative patient anatomy using the HoloLens 2},
  author={Kerkhof, Enzo and Thabit, Abdullah and Benmahdjoub, Mohamed and Ambrosini, Pierre and van Ginhoven, Tessa and Wolvius, Eppo B and van Walsum, Theo},
  journal={International Journal of Computer Assisted Radiology and Surgery},
  pages={1--12},
  year={2025},
  publisher={Springer}
}

@article{liu2024exploration,
  title={Exploration of the application of augmented reality technology for teaching spinal tumor’s anatomy and surgical techniques},
  author={Liu, Shuzhong and Yang, Jianxin and Jin, Hui and Liang, Annan and Zhang, Qi and Xing, Jinyi and Liu, Yong and Li, Shuangshou},
  journal={Frontiers in Medicine},
  pages={1403423},
  year={2024},
  publisher={Frontiers Media SA}
}

@article{jiang2023wearable,
  title={Wearable mechatronic ultrasound-integrated ar navigation system for lumbar puncture guidance},
  author={Jiang, Baichuan and Wang, Liam and Xu, Keshuai and Hossbach, Martin and Demir, Alican and Rajan, Purnima and Taylor, Russell H and Moghekar, Abhay and Foroughi, Pezhman and Kazanzides, Peter and others},
  journal={IEEE transactions on medical robotics and bionics},
  volume={5},
  number={4},
  pages={966--977},
  year={2023},
  publisher={IEEE}
}

@article{agten2018augmented,
  title={Augmented reality--guided lumbar facet joint injections},
  author={Agten, Christoph A and Dennler, Cyrill and Rosskopf, Andrea B and Jaberg, Laurenz and Pfirrmann, Christian WA and Farshad, Mazda},
  journal={Investigative radiology},
  volume={53},
  number={8},
  pages={495--498},
  year={2018},
  publisher={LWW}
}

@article{gibby2019head,
  title={Head-mounted display augmented reality to guide pedicle screw placement utilizing computed tomography},
  author={Gibby, Jacob T and Swenson, Samuel A and Cvetko, Steve and Rao, Raj and Javan, Ramin},
  journal={International journal of computer assisted radiology and surgery},
  volume={14},
  number={3},
  pages={525--535},
  year={2019},
  publisher={Springer}
}

@article{ma2025augmented,
  title={Augmented Reality Navigation System Enhances the Accuracy of Spinal Surgery Pedicle Screw Placement: A Randomized, Multicenter, Parallel-Controlled Clinical Trial},
  author={Ma, Yichao and Wu, Jiangpeng and Dong, Yanlong and Tang, Hongmei and Ma, Xiaojun},
  journal={Orthopaedic Surgery},
  volume={17},
  number={2},
  pages={631--643},
  year={2025},
  publisher={Wiley Online Library}
}

@article{altorfer2025pedicle,
  title={Pedicle screw placement with augmented reality versus Robotic-assisted surgery},
  author={Altorfer, Franziska CS and Kelly, Michael J and Avrumova, Fedan and Burkhard, Marco D and Zhu, Jiaqi and Abel, Frederik and Cammisa, Frank P and Sama, Andrew and Farshad, Mazda and Lebl, Darren R},
  journal={Spine},
  volume={50},
  number={15},
  pages={1058--1064},
  year={2025},
  publisher={LWW}
}

@article{karius2024first,
  title={First implementation of an innovative infra-red camera system integrated into a mobile CBCT scanner for applicator tracking in brachytherapy—Initial performance characterization},
  author={Karius, Andre and Leifeld, Lisa Marie and Strnad, Vratislav and Fietkau, Rainer and Bert, Christoph},
  journal={Journal of Applied Clinical Medical Physics},
  pages={e14364},
  year={2024},
  publisher={Wiley Online Library}
}

@article{jiang2021autonomous,
  title={Autonomous robotic screening of tubular structures based only on real-time ultrasound imaging feedback},
  author={Jiang, Zhongliang and Li, Zhenyu and Grimm, Matthias and Zhou, Mingchuan and Esposito, Marco and Wein, Wolfgang and Stechele, Walter and Wendler, Thomas and Navab, Nassir},
  journal={IEEE Transactions on Industrial Electronics},
  volume={69},
  number={7},
  pages={7064--7075},
  year={2021},
  publisher={IEEE}
}

@article{gorniak2015lower,
  title={Lower lumbar facet joint complex anatomy},
  author={Gorniak, G and Conrad, W},
  journal={Austin J Anat},
  volume={2},
  number={1},
  pages={1--8},
  year={2015}
}

@article{neumann1999determination,
  title={Determination of inter-spinous process distance in the lumbar spine: evaluation of reference population to facilitate detection of severe trauma},
  author={Neumann, P and Wang, Y and K{\"a}rrholm, J and Malchau, H and Nordwall, A},
  journal={European Spine Journal},
  volume={8},
  number={4},
  pages={272--278},
  year={1999},
  publisher={Springer}
}

@article{feng2020morphological,
  title={Morphological parameters of fourth lumbar spinous process palpation: a three-dimensional reconstruction of computed tomography},
  author={Feng, Qi and Zhang, Lei and Zhang, Mengyao and Wen, Youliang and Zhang, Ping and Wang, Yi and Zeng, Yan and Wang, Junqiu},
  journal={Journal of orthopaedic surgery and research},
  volume={15},
  number={1},
  pages={227},
  year={2020},
  publisher={Springer}
}

@article{farshad2020ultrasound,
  title={Ultrasound-guided interventions with augmented reality in situ visualisation: a proof-of-mechanism phantom study},
  author={Farshad-Amacker, Nadja A and Bay, Till and Rosskopf, Andrea B and Spirig, Jos{\'e} M and Wanivenhaus, Florian and Pfirrmann, Christian WA and Farshad, Mazda},
  journal={European radiology experimental},
  volume={4},
  number={1},
  pages={7},
  year={2020},
  publisher={Springer}
}

@article{von2022ultrarsound,
  title={UltrARsound: in situ visualization of live ultrasound images using HoloLens 2},
  author={Von Haxthausen, Felix and Moreta-Martinez, Rafael and Pose D{\'\i}ez de la Lastra, Alicia and Pascau, Javier and Ernst, Floris},
  journal={International Journal of Computer Assisted Radiology and Surgery},
  volume={17},
  number={11},
  pages={2081--2091},
  year={2022},
  publisher={Springer}
}

@article{fedorov20123d,
  title={3D Slicer as an image computing platform for the Quantitative Imaging Network},
  author={Fedorov, Andriy and Beichel, Reinhard and Kalpathy-Cramer, Jayashree and Finet, Julien and Fillion-Robin, Jean-Christophe and Pujol, Sonia and Bauer, Christian and Jennings, Dominique and Fennessy, Fiona and Sonka, Milan and others},
  journal={Magnetic resonance imaging},
  volume={30},
  number={9},
  pages={1323--1341},
  year={2012},
  publisher={Elsevier}
}

@article{li2026ultrasound,
  title={Ultrasound-Guided Real-Time Spinal Motion Visualization for Spinal Instability Assessment},
  author={Li, Feng and Bi, Yuan and Song, Tianyu and Jiang, Zhongliang and Navab, Nassir},
  journal={arXiv preprint arXiv:2602.12917},
  year={2026}
}

@article{song2026comparative,
  title={Comparative study of ultrasound shape completion and CBCT-based AR workflows for spinal needle interventions},
  author={Song, Tianyu and Li, Feng and Pabst, Felix and Gafencu, Miruna-Alexandra and Bi, Yuan and Eck, Ulrich and Navab, Nassir},
  journal={International Journal of Computer Assisted Radiology and Surgery},
  pages={1--9},
  year={2026},
  publisher={Springer}
}

@article{li2026robotic,
  title={Robotic Ultrasound Makes CBCT Alive},
  author={Li, Feng and Li, Ziyuan and Jiang, Zhongliang and Navab, Nassir and Bi, Yuan},
  journal={arXiv preprint arXiv:2603.10220},
  year={2026}
}

\end{document}